\documentclass{aa}  

\usepackage{graphicx}
\usepackage{natbib}
\usepackage{placeins}
\usepackage{xcolor} 

\bibpunct{(}{)}{;}{a}{}{,} 
\usepackage[varg]{txfonts}
\begin{document}

   \title{The colliding-wind binary HD~168112\thanks{Based on observations collected with
the Mercator Telescope operated on the island of La Palma by the
Flemish Community, at the Spanish Observatorio del Roque de los
Muchachos of the Instituto de Astrof\'{\i}sica de Canarias. Based on observations
obtained with the HERMES spectrograph, which is supported
by the Fund for Scientific Research of Flanders (FWO), Belgium,
the Research Council of K.U.Leuven, Belgium, the Fonds National
de la Recherche Scientifique (F.R.S.-FNRS), Belgium, the Royal
Observatory of Belgium, the Observatoire de Gen\`eve, Switzerland, and
the Th\"{u}ringer Landessternwarte Tautenburg, Germany. 
Also based on observations with the TIGRE telescope, located at La Luz observatory,
Mexico. TIGRE is a collaboration of the Hamburger Sternwarte,
the Universities of Hamburg, Guanajuato, and Liège.
Also based on observations obtained with the
Karl G. Jansky Very Large Array (VLA) of the
National Radio Astronomy 
Observatory (NRAO). The NRAO is a facility of the National Science Foundation 
operated under cooperative agreement by Associated Universities, Inc.
}}

   \subtitle{}

   \author{R. Blomme\inst{1}
          \and
          G. Rauw\inst{2}
          \and
          D. Volpi\inst{1,3}
          \and
          Y. Naz\'e\inst{2}\thanks{F.R.S.-FNRS Senior Research Associate}
          \and
          M. Abdul-Masih\inst{4,5}
          }

   \institute{Royal Observatory of Belgium,
              Ringlaan 3, B-1180 Brussels, Belgium\\
              \email{Ronny.Blomme@oma.be}
             \and
              Space sciences, Technologies and Astrophysics Research (STAR) Institute, Universit\'e de Li\`ege, All\'ee du Six Ao\^{u}t, 19c, B\^at B5c, 4000 Li\`ege, Belgium
             \and
              Facult\'e Science de la Motricit\'e, Universit\'e Libre de Bruxelles, Campus Erasme, Route de Lennik
              808, 1070 Anderlecht, Belgium
              \and
              Instituto de Astrof\'isica de Canarias, C. V\'ia L\'actea, s/n, 38205 La Laguna, Santa Cruz de Tenerife, Spain
              \and
              Universidad de La Laguna, Departamento de Astrof\'isica, Avenida Astrof\'isico Francisco S\'anchez s/n, 38206 La Laguna, Tenerife, Spain
             }

   \date{Received <date>; accepted <date>}

 
  \abstract
   {Radio surveys of early-type stars have revealed a number of non-thermal emitters. Most of these have been shown 
   	to be binaries, where the collision between the two stellar winds is responsible for the non-thermal emission. 
   }
   {HD~168112 is a non-thermal radio emitter, whose binary nature has only recently been 
   	confirmed spectroscopically. We obtained independent spectroscopic observations to determine its orbit, in addition to
   	radio observations to see
   	if the thermal or non-thermal nature of the emission changes during the periastron passage. 
   }
   {We monitored HD~168112 spectroscopically for a 13 year time span. 
   	From these data, we determined the orbital parameters, which we compared to the previous results in the literature.
   	The stellar parameters of both components
   	were determined by comparing the spectra to TLUSTY models. 
   	From the spectral index of the radio observations, we found how the nature of the
   	emission changes as the system goes through periastron. Combining our results
   	with other literature data allowed us to further constrain the orbital and stellar parameters.
   }
   {We find HD 168112 to have an orbital period of $P = 512.17^{+0.41}_{-0.11}$~d, an eccentricity of $e = 0.7533^{+0.0053}_{-0.0124}$,
   	and a mass ratio close to one. From our spectroscopic modelling, we derived the stellar parameters, but 
   	we had difficulty arriving at a spectroscopic mass ratio of one. The radio observations
   	around periastron
    show only thermal emission, suggesting that most of the synchrotron photons are absorbed
   	in the two stellar winds at that phase. Combining our data with the optical interferometry detection, 
   	we could constrain the inclination angle to $i \sim 63\degr$, and the 
   	mass of each component to $\sim 26$ M$_\sun$.    	
   }
   {We have provided an independent spectroscopic confirmation of the binary nature of
   	HD~168112. Although detected as a non-thermal radio emitter, near periastron the radio emission
   	of this highly eccentric system is thermal and is mainly formed in the colliding-wind region. 
   	This effect will also occur in other colliding-wind binaries.
   }

   \maketitle
%

\section{Introduction}

The first survey of radio emission in hot stars was done by \cite{Bieging+89}.
Among these stars, a good number of thermal radio emitters were detected,
where the emission is due to free-free processes in the ionized stellar wind.
The measured radio flux of these stars can be used to determine the mass-loss rate
\citep{Wright+Barlow75, Panagia+Felli75}. 

However, the survey also unveiled a number of non-thermal radio emitters, as recognized
by the spectral index of their 
emission\footnote{The spectral index is the quantity $\alpha$ in $F_\nu \propto \nu^\alpha$.
        Thermal emission in the stellar wind gives $\alpha \approx +0.6$. Non-thermal
        emitters have $\alpha \le 0.0$.
        }.
They also have
an excess flux compared to that expected from the free-free wind
emission, a high brightness temperature, and a variable flux.
Since that first survey, substantially more non-thermal radio emitters
have been found \citep[for a review and a catalogue, see e.g.][]{DeBeckerReview07, DeBecker+Raucq13}.

For Wolf-Rayet stars, it quickly became clear that 
the non-thermal emission is mostly due to colliding-wind binaries \citep{Dougherty+Williams00}. 
In a massive-star binary system, the two stellar winds collide and shocks form on either side of the collision
region. Around these shocks, a fraction of electrons can be accelerated by the Fermi mechanism 
to relativistic speeds \citep{Bell78,Reitberger+14, Pittard+21}. By spiralling around in the magnetic
field, these electrons then emit synchrotron radiation, which we detect as non-thermal emission 
\citep{Eichler+Usov93}.

The situation for the O-type non-thermal emitters was not so clear. 
A number of sources were known to be binaries:
\object{HD~15558} \citep[][and references therein]{DeBecker+06},
\object{HD~167971} \citep[a triple system - ][and references therein]{Leitherer+87}.
Over time,
most of the other sources classified by \cite{Bieging+89} as `definite' or `probable'
non-thermal emitters were also found to be binaries:
\object{9~Sgr} \citep{Rauw+16},
\object{Cyg~OB2~\#8A} \citep{DeBecker+04-8A}, and
\object{Cyg~OB2~\#9} \citep{Naze+08}.

\object{HD~168112} was
also listed as a non-thermal radio emitter by \cite{Bieging+89}, but for a long time
there was only indirect evidence of it being a binary.
\cite{DeBecker+04-168112} confirmed the non-thermal radio emission and showed that the X-ray 
emission is variable, but their optical spectra did not show any indication of binarity.
\cite{Blomme+05} found periodic behaviour in the archive radio data, which 
does suggest that it is a binary system.
A subsequent X-ray Multi-Mirror Mission (XMM-Newton) observation by \cite{DeBecker15} showed the X-ray emission to be overluminous, which can
be attributed to heated material in the colliding-wind region.
\cite{Sana+14} obtained optical interferometry observations
with the Precision Integrated-Optics Near-infrared Imaging ExpeRiment (PIONIER) instrument at the Very Large Telescope 
Interferometer (VLTI) and found a companion at an angular
distance of $3.33 \pm 0.17$~mas with a brightness difference
in the H-band of $0.17 \pm 0.19$~mag.

Early attempts by \cite{Rauw+05} and \cite{Chini+12} to find spectroscopic evidence for binarity
were not successful. However, \cite{MaizApellaniz+19} succeeded in assigning two spectral
types to the components of HD~168112: O5 IV(f) + O6: IV: or   O4.5 III(f) + O5.5 IV((f))
for the LiLiMaRlin data \citep[Library of Libraries of Massive-Star High-Resolution Spectra,][]{lillimarlin}.

Finally, \cite{Putkuri+23} published a spectroscopic orbit determination, providing convincing evidence
that HD~168112 is indeed a binary.
They found it to be a highly eccentric system (with eccentricity $e = 0.743 \pm 0.005$) and a period
of $P = 513.52 \pm 0.01$~d. The masses of the two components are very similar and the minimum value
they derived is $\sim~27.1~{\rm M}_\sun$ for the A component and $\sim~24.6~{\rm M}_\sun$ for the B component.

Additionally, radio observations by \cite{DeBecker+24} resolved the colliding-wind region of \object{HD~168112}. They used 
the European Very Long Baseline Interferometry Network (EVN) to observe at 1.6 GHz ($18$~cm).
The colliding-wind region
is clearly asymmetric and its high flux (integrated $1.7\pm0.4$~mJy) shows it to be non-thermal in origin, thereby
providing further evidence for the colliding-wind nature of the HD 168112 binary. 

Colliding-wind binaries are highly interesting because they provide the opportunity to study the acceleration of
particles around shocks. This process occurs in a number of astrophysical environments, such as interplanetary shocks
and supernova remnants. The colliding-wind binaries provide an environment that is quite different from the others
regarding parameters such as
density, magnetic field, ambient radiation field, and shock speed. They are also relevant for the
mass-loss rate determinations in single stars, as they can provide an independent
determination of the effect of clumping and porosity \citep[e.g.][]{Pittard07}. This in turn
will help determine the stellar mass-loss rates, which are very sensitive to clumping and porosity
\citep[e.g.][]{Puls+08}.

In this paper, we present independent spectroscopic observations of HD 168112, covering a time span of 13 years.
From these data, we derived an orbital solution. The combination of the spectroscopic data and the optical interferometry
results of \cite{Sana+14} allowed us to further constrain the orbital parameters. 
We also determined the stellar parameters of both components.
Furthermore, in this work, we analyse the radio observations we obtained 
near the 2013 periastron passage of HD~168112. These observations are part of a coordinated XMM-Newton and VLA project.
The XMM-Newton data are analysed in a separate paper \citep{Rauw+24}.

In Sect.~\ref{section data}, we present both the spectroscopic data and radio data we obtained. 
Sect.~\ref{section spectroscopic data} analyses the spectroscopic data, determining the orbital parameters
and the stellar parameters of both components of the binary system. In Sect.~\ref{section radio data}
we analyse the new radio data. Sect.~\ref{section other data} combines our results with data from the literature
to further determine the orbital and stellar parameters. Sect.~\ref{section conclusions} presents our conclusions.

\section{Data}
\label{section data}

\subsection{Optical spectroscopy}

\object{HD~168112} was monitored during 13 years with the HERMES \cite[High Efficiency and 
Resolution Mercator Echelle Spectrograph,][]{Raskin+11} spectrograph on 
the 1.2~m Mercator telescope at the Roque de los Muchachos Observatory (La Palma, 
Canary Islands, Spain). 
We obtained 97
usable\footnote{HERMES spectra numbers 415762, 415763, 578768 were
       not usable as they contain only noise.}
HERMES spectra between June 2009 and September 2021.
Our HERMES spectra are part of another observing programme than those used
by \cite{Putkuri+23}, and therefore provide an independent set of data.

The HERMES instrument has a resolving power R $\approx$ 85\,000 and covers the 
wavelength range 3770 to 9000 $\AA$.
The data were reduced using the standard HERMES 
pipeline\footnote{\url{http://hermes-as.oma.be/}}.
Spectral orders were extracted and flat-fielding and wavelength calibration 
(based on Th-Ar lamp spectra) was applied. Cosmic rays were removed, the orders 
were merged, and the barycentric velocity correction was applied.

An additional 16 spectra were obtained with the refurbished HEROS 
(Heidelberg Extended Range Optical Spectrograph) instrument
on the 1.2~m Telescopio Internacional de Guanajuato Rob\'{o}tico
Espectrosc\'{o}pico \citep[TIGRE,][]{TIGRE,TIGRE2}, located at La Luz observatory,
Mexico. The resolving power is 20\,000, covering the wavelength range 3800 -- 8800 $\AA$.
Data reduced with version 3.1 of the pipeline \citep{TIGRE-DataReduction}
were downloaded from the TIGRE archive\footnote{\url{https://hsweb.hs.uni-hamburg.de/projects/TIGRE/EN/Archive/login.php}}.
We applied the barycentric correction to the spectra.
The observing log of both the HERMES and TIGRE observations is given in
Table~\ref{table:obslog}.

\subsection{Radio data}

We obtained radio observations of \object{HD~168112} near the periastron passage, using the Karl G. Jansky Very Large Array (VLA) of the
National Radio Astronomy 
Observatory (NRAO).
The VLA project SX222004 (PI: GR) is the VLA part of a joint 
XMM-Newton and NRAO proposal.
The X-ray data are presented in a separate paper \citep{Rauw+24}.
We observed on three dates: one near periastron (2023-Mar-18) and the others approximately two weeks before 
(Mar-03) and after periastron (Apr-05). For all three observations, the VLA was
in B configuration.

Data were obtained at 3.6 cm (X-band) and 6 cm (C-band)
On each date and for each band, three sequences of observations were done, consisting of a $\sim$~4.6 minute observation on \object{HD~168112},
preceded and followed by a  $\sim$~1.5 minute observation on the phase calibrator \object{J1832-1035} (which is 3.7\degr\ from \object{HD~168112}). 
The total observing time on \object{HD~168112} is therefore $\sim$~14 minutes, for each date and each band.
At the end of the run, the flux calibrator \object{3C286} was observed ($\sim$~1 minute for C-band,  $\sim$~1.5 minute for X-band). 
The data were taken in 32 spectral windows of 128 MHz each, with the X-band covering 8.0--12.0 GHz and the C-band
4.0--8.0 GHz. Each spectral window in turn consists of 64 channels of 2 MHz each.

\begin{figure}[]
\resizebox{\hsize}{!}{\includegraphics[bb=51 37 366 302,clip]{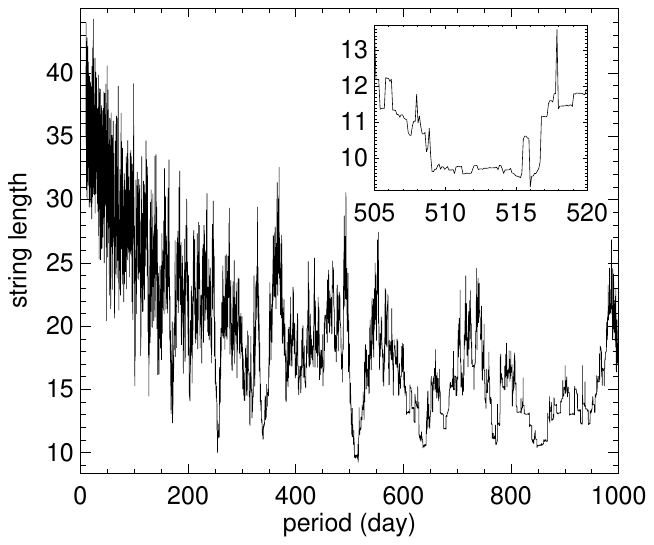}}
\caption{String length as a function of the period.
The inset shows a zoom-in on the best value found.
}
\label{fig string length}
\end{figure}

\section{Analysis of spectroscopic data}
\label{section spectroscopic data}

Consistent with its O5 III spectral classification \cite[as a single star, ][]{Holgado+18}, the spectra of \object{HD~168112}
show prominent hydrogen, \ion{He}{I} and \ion{He}{II} lines, as well as some metal lines.
The hydrogen and helium lines do not show P Cygni profiles, indicating a
relatively weak stellar wind. 
The \ion{C}{III} 5696, 6730, and \ion{N}{III} 4634,40,41 lines are in emission.

While most of the spectra are single-lined, some of them show clearly 
double-lined features, indicating the binary nature of
\object{HD~168112}. 
Some of the apparently single-lined spectra show line broadening, which we attribute
to radial velocity differences which are too small to induce a double-lined
profile. 

In our analysis, we applied a spectral disentangling technique to the data (details are given in Sect.~\ref{section disentangling}).
To start up the disentangling procedure, we obtained a first estimate of the
radial velocities and a first estimate of the orbital parameters (Sect.~\ref{section preliminary orbit}).
We then used the disentangled spectra to update the radial velocities and redetermined the orbit.
Iterating this a number of times led to the final orbital parameters (Sect.~\ref{section final orbit}).
From the disentangled spectra, we then found the rotational and macroturbulent velocities (Sect.~\ref{section vsini}).
By comparison to TLUSTY models, we then determined the stellar parameters (Sect.~\ref{section stellar parameters}).

\subsection{Preliminary orbit}
\label{section preliminary orbit}

We determined preliminary radial velocities by fitting a selected set of
spectral lines with one or multiple Gaussians.
While there are a good number of spectral lines that show the line-doubling, we limited this analysis to 
those with a high enough signal-to-noise ratio (S/N).
We therefore first determined the S/N around a large number of candidate lines.
The S/N at 5410 \AA\ is listed in Table~\ref{table:obslog}, and the variation of S/N across the spectra is shown
in Fig.~\ref{fig S/N}.
On the basis of this, we chose to continue with the following spectral lines:
\ion{He}{I} 4471,
\ion{He}{I} 5876,
\ion{He}{II} 4200,
\ion{He}{II} 4541,
\ion{He}{II} 5411,
\ion{C}{IV} 5801,12, and
\ion{O}{III} 5592.

In the region around each of these lines, we flagged any remaining cosmic rays. We then normalised the spectra on each of these regions
by fitting a second-degree polynomial through interactively selected continuum points.
Where the line is clearly split, we obtained two radial velocities (RVs); otherwise only one value was measured.
For each spectrum, we then took an average of the individual RVs, weighting them with the inverse of the square of
the individual error bars. In this process, we excluded any outlier RVs.

\begin{figure*}
\sidecaption
\includegraphics[width=12cm,bb=40  5 780 525,clip]{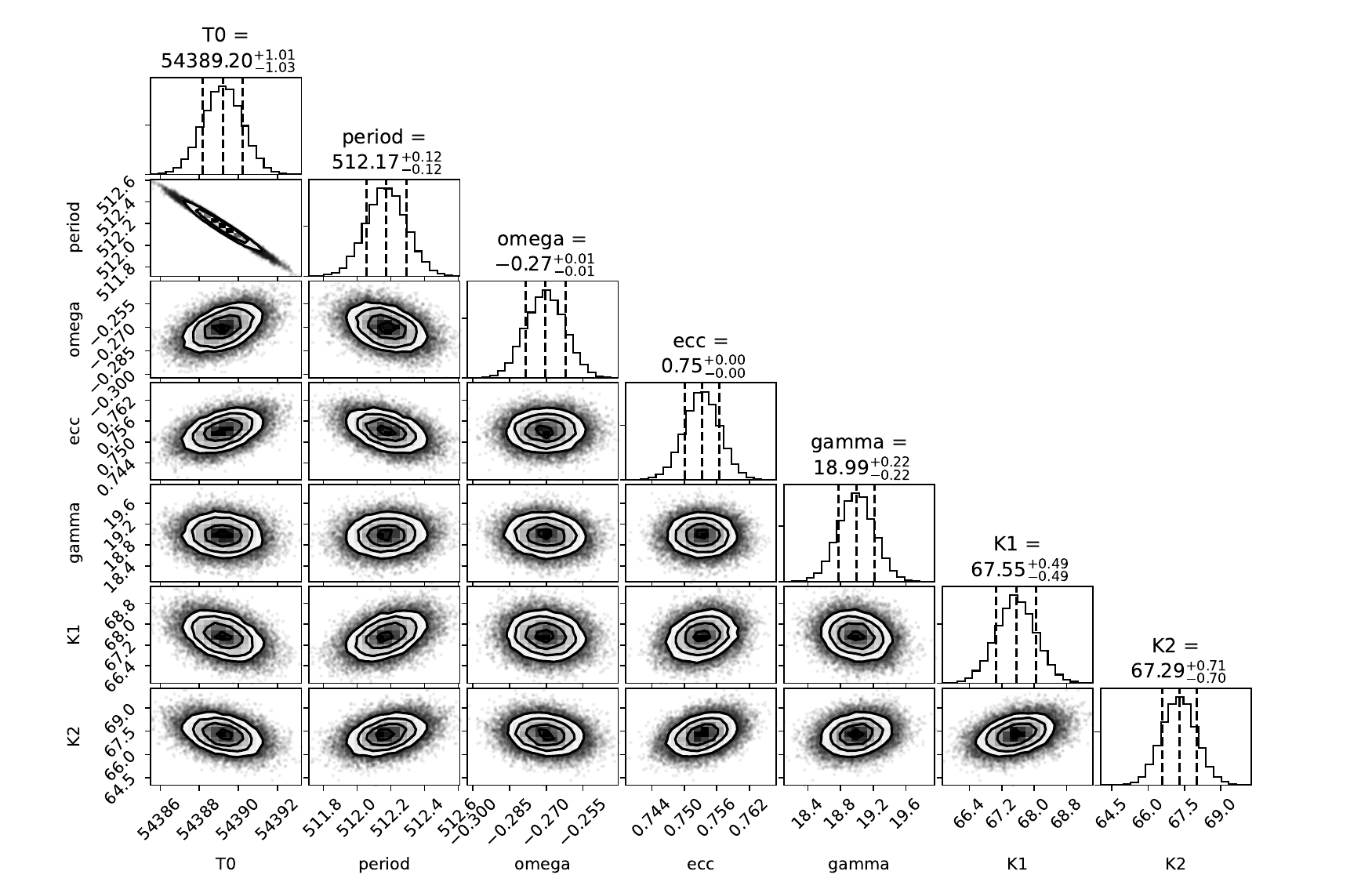}
\caption{Corner plot showing the final orbital parameters and their correlations.
                The time of the periastron passage ($T_0$) and the period are
                strongly correlated.}
\label{corner plot}
\end{figure*}

To determine the orbital period from these RV data, we applied 
the string length method \citep{stringlength}.
We rescaled
the radial velocities of each component separately to the range 0..1.
This ensured that the phase differences and the radial velocity differences
have equal weight.
Figure~\ref{fig string length} shows that 
the best result for the period is $\sim 516$ days.
The second lowest value is at $\sim 255$ days, which is close to half of the best period.
We checked the spectra taken near the additional periastron passages of an assumed $\sim 255$ day period.
Only data taken around JD 2\,457\,200 fulfil this criterion, and they clearly show no splitting of the spectral lines.
We therefore took $\sim 516$ days as the (approximate) value for the period.

We then determined the preliminary orbit based on these RVs, using the Markov Chain Monte Carlo (MCMC) code \textit{emcee} \citep{emcee}.
This code explores many combinations of the orbital parameters
$T_0$ (time of periastron passage),
$P$ (period),
$e$ (eccentricity),
$\omega$ (longitude of periastron),
$\gamma$ (systemic 
velocity\footnote{Sometimes it is necessary to assign different values for the systemic velocity of the two components
	\citep{Rauw+02}. 
	We ran experiments with two $\gamma$ values instead of one,
	but the final orbital parameters showed the two values to be very close to each other.}), and
$K_1$, $K_2$ (semi-amplitude radial velocity of primary and secondary).
For each combination of these parameters the code calculates the model radial velocities 
(${\rm RVModel}_{1,i}$, ${\rm RVModel}_{2,i}$) for the
times of the observed spectra.
The code then finds the solution with the smallest $\chi^2$, where:
\begin{equation}
\chi^2 =\sum_i \left( \frac{{\rm RVObs}_{1,i}-{\rm RVModel}_{1,i}}{{\rm RVObsError}_{1,i}} \right)^2 + \left( \frac{{\rm RVObs}_{2,i}-{\rm RVModel}_{2,i}}{{\rm RVObsError}_{2,i}} \right)^2, \nonumber
\end{equation}
where ${\rm RVObs}_{1,i}$ and ${\rm RVObs}_{2,i}$ are the observed radial velocities of the first and second component,
with their associated error bars ${\rm RVObsError}_{1,i}$ and ${\rm RVObsError}_{2,i}$.
For those spectra where we could not split the lines, we used ${\rm RVObs}_{2,i}={\rm RVObs}_{1,i}$.

In the MCMC procedure we used no significant priors, except requiring an eccentricity between
0 and 1. We used 32 walkers and started them with a range in value that covers the above derived approximate period,
as well as estimates of the semi-amplitude radial velocities.
We ran it for 20\,000 iterations and discarded
the first 5000. Based on estimates of the auto-correlation times,
we thinned the resulting samples by taking only one in every 200.

\subsection{Disentangling}
\label{section disentangling}

Using the orbital parameters derived above, we used a disentangling technique on each of the spectral lines.
For the disentangling, we used the code by \cite{Shenar+20, Shenar+22}. This code is based on the shift and add technique
introduced by \cite{Marchenko+98} and \cite{Gonzalez+Levato}. At each step in this iterative procedure, an approximate spectrum is available
for each of the two binary components. The next step consists of subtracting the Doppler-shifted intrinsic spectrum of the first
component from the observed spectra, thus giving spectra that have only contributions of the second component. Doppler-shifting
these spectra and adding them up gives the next approximation of the intrinsic spectrum of the second component. The procedure
is then applied again, exchanging the roles of the first and second component. After a sufficient number of iterations,
the disentangled spectra of the two binary components are obtained.

We applied the disentangling part of the \cite{Shenar+20, Shenar+22} code to each of the spectral lines we used in the 
radial velocity determination (Sect.~\ref{section preliminary orbit}). We applied 1000 iterations; checks with a
higher number of iterations gave the same result.
For each of these lines we thus obtained two separate lines, one for each component of the binary.
We then explored for each spectrum and each spectral line a two-dimensional grid of velocities. For every point in the
grid we shifted the two disentangled spectral lines by the two chosen velocities and added them together. This was then compared 
to the observed spectral line at that epoch, and the $\chi^2$ difference was determined. From the minimum $\chi^2$, the 
RVs of the two components at that epoch were then derived. In this procedure, we assumed that the light ratio between the
two components is 1.0; the correct value is determined in Sect.~\ref{section stellar parameters}.

The combined RVs for each epoch were then determined as the equal-weight average of the RVs of the individual spectral lines.
Outliers were removed in this procedure. The error bar on the combined RVs is given by the standard deviation on the 
individual RV values that contributed to the combined one.

\subsection{Final orbital parameters}
\label{section final orbit}

To determine the final orbital parameters, we iterated between the disentangling and the orbital parameter determination
outlined in the previous
sections. At each step in the iteration, we used the radial velocities as input to the \textit{emcee} code, which determined
the orbital parameters. From these orbital parameters, the disentangling code found the spectral lines of the two
binary components. These were then used to redetermine the radial velocities. 

\begin{figure}[]
\centering
\resizebox{\hsize}{!}{\includegraphics[viewport=33 0 365 740,clip]{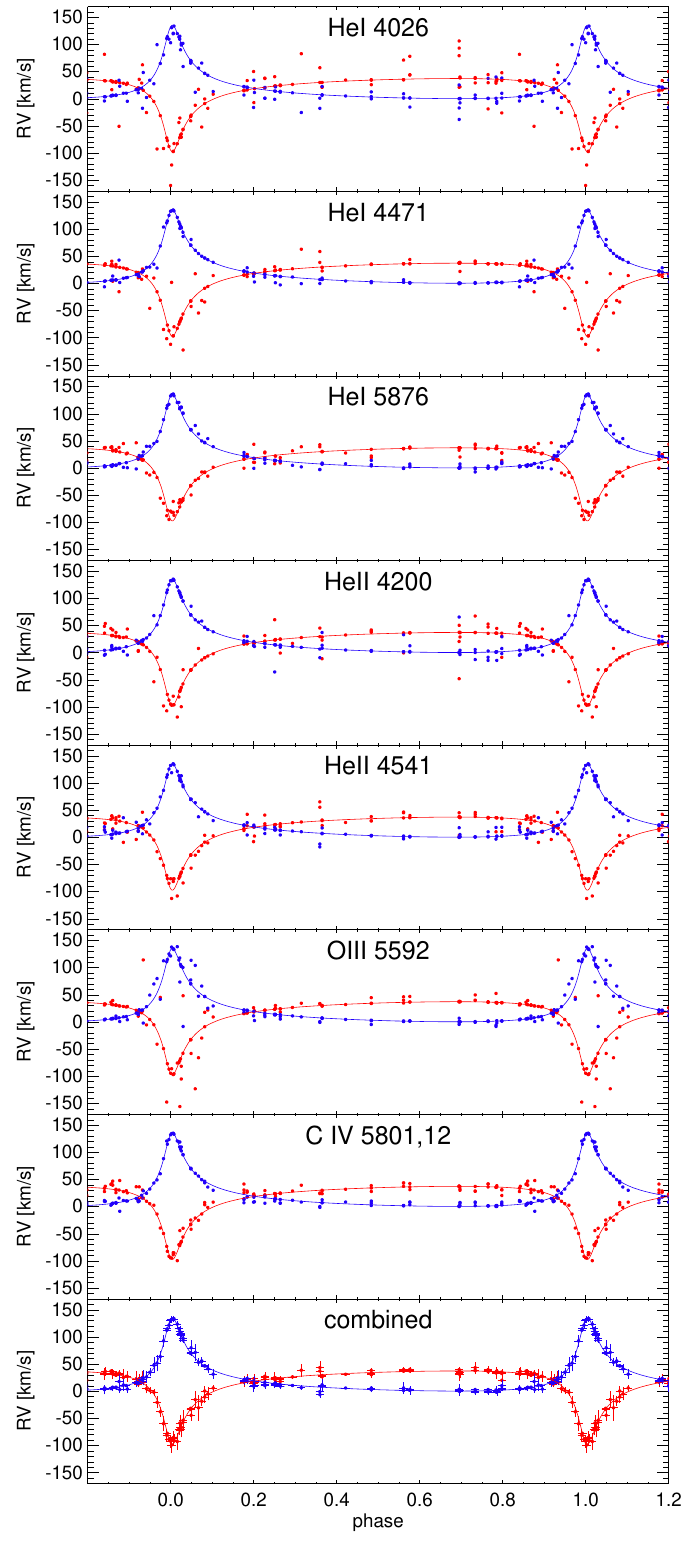}}
\caption{
Comparison of our orbital solution (solid line) to the observed radial velocities (symbols). Each panel shows the observed values for
a specific stellar line, with the bottom panel showing the radial velocities combined over all spectral lines. Blue indicates
the A component, red the B component.}
\label{fig RVs}
\end{figure}

During these iterations we checked the convergence of the orbital parameters. We stopped after 20 iterations,
as the parameters had then sufficiently converged. To obtain the final values, we ran a longer \textit{emcee} run
with 200\,000 iterations, discarding the first 50\,000. The corner plot presenting the results of this
\textit{emcee} run is shown in Fig.~\ref{corner plot}. 
The fit of our orbital solution to the observed radial velocities, for each spectral line, and for their combination,
is shown in Fig.~\ref{fig RVs}.
The orbital parameters are listed in Table~\ref{table orbital parameters}.
As noted in Table~\ref{table orbital parameters}, the mass ratio is very close to one, making it difficult to decide which of
the two components is the primary. To avoid introducing confusion in the literature,
 we have assigned the subscript ``A'' to the component that \cite{Putkuri+23} designate as the primary, and the subscript ``B'' to the secondary.

The MCMC approach is very good at determining the error bar on the orbital parameters due to the error bars in the
radial velocities. But we also wanted to get an estimate of any systematic errors. We therefore ran the whole
iterative procedure again with some changes to the input. We experimented with the wavelength range covered by each of 
the spectral lines, as well as systematically dropping one of the spectral lines from the procedure.
We extended the range of the MCMC error bars to also include the values covered by these variant analyses.
The final error bars are listed in Table~\ref{table orbital parameters}.

\subsection{Comparison with \cite{Putkuri+23}}

\begin{figure}
	\resizebox{\hsize}{!}{\includegraphics[bb=30 35 325 310,clip]{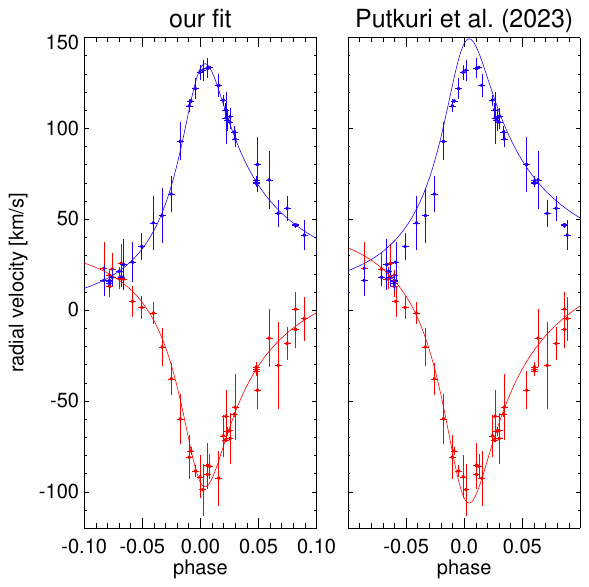}}
	\caption{Comparison of our orbital solution to the \cite{Putkuri+23} one 
		for our data around periastron. The phases on the left panel are calculated with
		our ephemeris, the ones on the right panel with the \cite{Putkuri+23} one.
		The symbols show our observed radial velocities (combined data), on both panels. The colour coding
		is as in Fig.~\ref{fig RVs}. The solid lines indicate the orbital solutions for the respective components.
	}
	\label{fig comparison RV}
\end{figure}

\begin{table}
	\caption{Orbital parameters of HD~168112.}
	\label{table orbital parameters}
	\begin{center}
	\begin{tabular}{lc}
		\hline\hline
		parameter & value \\
		\hline		
		\noalign{\smallskip}
		$T_0$ (BJD-2\,400\,000)    & $54389.20_{-3.58}^{+1.01}$ \\
		\noalign{\smallskip}
		$P$ (day)                  & $512.17_{-0.11}^{+0.41}$ \\
		\noalign{\smallskip}
		$e$                        & $0.7533_{-0.0124}^{+0.0053}$ \\
		\noalign{\smallskip}
		$\omega_A$ (deg)           & $344.51_{-0.43}^{+1.54}$ \\
		\noalign{\smallskip}
        $\gamma$ (km\,s$^{-1}$)    & $18.99_{-1.88}^{+0.22}$ \\
        \noalign{\smallskip}
        $K_A$ (km\,s$^{-1}$)       & $67.55_{-2.34}^{+0.55}$ \\
        \noalign{\smallskip}
       	$K_B$ (km\,s$^{-1}$)       & $67.29_{-5.13}^{+0.80}$ \\
       	\noalign{\smallskip}
       	$a_A \sin i$ (R$_\sun$)    & $449.6_{-17.6}^{+19.8}$ \\
        \noalign{\smallskip}
        $a_B \sin i$ (R$_\sun$)   & $447.8_{-35.1}^{+20.1}$  \\
        \noalign{\smallskip}
        $M_A/M_B$                  & $0.996_{-0.183}^{+0.105}$ \\
        \noalign{\smallskip}
        $M_A \sin^3 i$ (M$_\sun$)  & $18.5_{-3.2}^{+1.5}$ \\
        \noalign{\smallskip}
        $M_B \sin^3 i$ (M$_\sun$)  & $18.6_{-1.7}^{+1.3}$ \\
        \noalign{\smallskip}
		\hline		
	\end{tabular}
	\end{center}
\tablefoot{
The masses of both stars are equal within their error bars, so it is not possible from our data
to decide which one is the primary. To avoid introducing confusion in the literature,
 we have assigned the subscript ``A'' to the component that \cite{Putkuri+23} designate as the primary.
}
\end{table}

A comparison with the \cite{Putkuri+23} orbital determination shows that they have a slightly longer period ($513.52$~d) than our value of
$512.17$~d. Converting their $T_0$ to a value we can compare with our Fig.~\ref{corner plot}, we find $54\,377.39$. This combination
of $T_0$ and period is well outside the corresponding plot in Fig.~\ref{corner plot}. The \cite{Putkuri+23} period has an
unexpectedly small error bar: 0.01 d. Accumulated over the $\sim$ 11 cycles covered by the time span of their observations, this would
result in a phase difference of $2\times 10^{-4}$, which would be hard to detect. Our $\sim$ 0.4 d error bar over $\sim$ 9 cycles
of the time span covered by our observations results in a phase difference of $7 \times 10^{-3}$, which is just about
detectable around periastron. 

When we plot our radial velocities in
an orbital phase diagram using the \cite{Putkuri+23} orbital solution (Fig.~\ref{fig comparison RV}), 
we first note that our measured radial velocities are somewhat smaller than the \cite{Putkuri+23} orbital fit suggests.
This difference is due to the different way we measure the radial velocities. \cite{Putkuri+23} fitted Gaussian profiles
to the spectral lines, while we applied an iterative procedure using profiles from the disentangled spectra,
as detailed in the above section.

The orbital fit to our data using the \cite{Putkuri+23} parameters has a 
peak at periastron that is less sharp.
It is important to note that around periastron we have a higher number of observations than \citet[their Fig.~2]{Putkuri+23},
which allows us to better estimate $T_0$ and the period. 
It also allows us to derive a better constraint on the radial velocity semi-amplitudes, which
turn out to be somewhat smaller than their results
(which are $K_A = 70.4 \pm 1.1$ km\,s$^{-1}$, $K_B=77.5 \pm 1.2$ km\,s$^{-1}$).
The smaller semi-amplitudes also explain the differences in the  derived parameters of minimum semi-major axis,
minimum masses, and mass ratio.

Our value for the eccentricity is slightly higher than the \cite{Putkuri+23} one ($e=0.743 \pm 0.005$), but the error bars overlap.
The values for the longitude of periastron are equal within the error bars. 

\subsection{Projected rotational velocity}
\label{section vsini}

To determine the projected rotational velocity ($v \sin i$), we used the 
{\sc iacob-broad} code \citep{IACOB_BROAD}. This code determines both the
$v \sin i$ and the macroturbulent broadening ($v_{\rm mac}$). One part of the code
uses the Fourier transform of the spectral line, and from the first
zero in the transform finds $v \sin i$ \citep{Simon-Diaz+Herrero07}.
Another part of the code explores a number of $v \sin i$ and
$v_{\rm mac}$ combinations, calculates the resulting line profile,
and uses a goodness-of-fit technique ($\chi^2$ minimization) 
to determine the best-fit values.

Stark-broadened lines are not suitable for the application of this technique. We therefore 
limited it to the two metallic lines that have a sufficiently high signal-to-noise
in our disentangled spectra: \ion{O}{III} 5592 and \ion{C}{IV} 5812 (the \ion{C}{IV} 5801 line is not
well disentangled due to a nearby Diffuse Interstellar Band). 

For the A component, the \ion{O}{III} 5592 line gave
$v \sin i_{\rm fou}=81$~km\,s$^{-1}$ from the Fourier technique
and $v \sin i_{\rm gof}=82_{-32}^{+17}$~km\,s$^{-1}$ and $v_{\rm mac}=48_{-47}^{+47}$~km\,s$^{-1}$ from the
goodness-of-fit. The \ion{C}{IV} 5812 line gave, respectively,
$v \sin i_{\rm fou}=94$~km\,s$^{-1}$,
$v \sin i_{\rm gof}=86_{-23}^{+14}$~km\,s$^{-1}$,
and $v_{\rm mac}=72_{-28}^{+31}$~km\,s$^{-1}$.
From the range covered by these determinations, we 
assigned $v \sin i=75 \pm 25$~km\,s$^{-1}$ to the primary component.
We could not determine a reliable value for the macroturbulent broadening, only 
an upper limit of $v_{\rm mac}$ of $\sim 100$~km\,s$^{-1}$.

For the B component, we found 
$v \sin i_{\rm fou}=141$~km\,s$^{-1}$,
$v \sin i_{\rm gof}=141_{-23}^{+13}$~km\,s$^{-1}$,
and $v_{\rm mac}=64_{-44}^{+52}$~km\,s$^{-1}$  from \ion{O}{III} 5592, and
$v \sin i_{\rm fou}=129$~km\,s$^{-1}$,
$v \sin i_{\rm gof}=130_{-17}^{+12}$~km\,s$^{-1}$,
and $v_{\rm mac}=57_{-46}^{+39}$~km\,s$^{-1}$ from \ion{C}{IV} 5812.
From the range covered by these determinations, we combined this to $v \sin i=134 \pm 21$~km\,s$^{-1}$ and an upper limit
for $v_{\rm mac}$ of  $\sim 110$~km\,s$^{-1}$ for the B component.

\subsection{Stellar parameters}
\label{section stellar parameters}

To determine the stellar parameters, we first created a grid of TLUSTY models.
We started from the models with solar composition in the public OSTAR2002 grid \citep{OSTAR_2002}.
We used the TLUSTY 
code\footnote{TLUSTY version 205 and SYNSPEC version 51 downloaded from \url{http://www.as.arizona.edu/~hubeny/pub/tlusty205.tar.gz};
	line list from \url{https://www.as.arizona.edu/~hubeny/tlusty208-package/linelist.tar.gz}.}
to refine this grid, covering the $T_\mathrm{eff}$ range of $30\,000-40\,000$~K with a step of 1000 K
and the $\log g=3.5-4.0$ range with a step of 0.1. We also covered the range $T_\mathrm{eff} = 40\,000-46\,000$~K with a step of 1000 K, 
$\log g=4.0-4.7$ with a step of 0.1. We used a helium abundance of 0.10 and a microturbulent velocity of 10 km\,s$^{-1}$.

To calculate the model spectrum, we used SYNSPEC \citep{SYNSPEC}. These synthetic spectra were then convolved with a
rotational profile and a macroturbulent profile, using a code similar to {\sc iacob-broad}. For the rotational broadening
of the A component, we used $v \sin i = 75$~km\,s$^{-1}$. As shown in the previous section, we could  derive only
an upper limit for the macroturbulent velocity, and we therefore took the \cite{Putkuri+23} value of
$v_{\rm mac} = 75$~km\,s$^{-1}$.
For the B component, we used our value of $v \sin i = 134$~km\,s$^{-1}$ and the \cite{Putkuri+23} value of 
$v_{\rm mac} = 100$~km\,s$^{-1}$.
Finally we applied the instrumental broadening, using a Gaussian with the resolving power of the instrument.

Contrary to the procedure of \cite{Putkuri+23}, we did not fit the disentangled spectra separately, but we compared combined 
synthetic spectra to the observed data. This has the advantage that it avoids the potential artefacts
associated with disentangling. To reduce the amount of data we had to process, we selected spectra for a number of representative
orbital phases (Table~\ref{table selected spectra}).
We gave a higher preference to phases around periastron. As the spectral lines are clearly split at these phases, 
they allow for a better constraint on the stellar parameters of the two components.

\begin{table}
	\caption{Selected spectra used in determining the stellar parameters.}
	\label{table selected spectra}
	\begin{center}
		\begin{tabular}{ll|ll}
			\hline\hline
			\multicolumn{1}{c}{phase} & \multicolumn{1}{c}{file} & \multicolumn{1}{c}{phase} & \multicolumn{1}{c}{file} \\
			\hline
0.006 & 00834212  & 0.421 & 00415761 \\
0.025 & 00835601  & 0.561 & 00651693\\
0.059 & T20170907 & 0.922 & 00288715\\
0.082 & 00844340  & 0.959 & T20200508\\
0.176 & 00717344  & 0.992 & 0096384\\
0.251 & 00724589  & \\
			\hline
		\end{tabular}
	\end{center}
\end{table}

\begin{table}
	\caption{Stellar parameters of HD~168112 A and B.}
	\label{table stellar parameters}
	\begin{center}
		\begin{tabular}{lcc}
			\hline\hline
			parameter & \multicolumn{2}{c}{value} \\
			& component A & component B \\
			\hline		
			\noalign{\smallskip}
			T$_{\rm eff}$ (K) & $41\,000\pm1000$ & $37\,000\pm1000$ \\
			$\log g$ (cgs) & $3.9\pm0.1$ & $3.45\pm0.1$ \\
			$M_V$ (mag)    & $-5.580\pm0.104$ & $-5.362\pm0.110$ \\
			Bol. corr. (mag) & $-3.94$ & $-3.67$ \\
			$\log L/L_\sun$ & $5.71\pm0.04$ & $5.51\pm0.04$ \\
			Radius (R$_\sun$) & $14.1\pm1.0$ & $13.9\pm1.0$ \\
			Mass (M$_\sun$) & $58\pm16$ & $23\pm6$ \\
			\hline		
		\end{tabular}
	\end{center}
\end{table}

We used the following spectral lines for fitting:
H$\gamma$,
\ion{He}{I} 4026,
\ion{He}{I} 4471,
\ion{He}{I} 5876,
\ion{He}{II} 4200,
\ion{He}{II} 4541, and
\ion{He}{II} 5411.
We looped over all combinations of a theoretical spectrum for the A component and one for the B
component. We applied the required radial velocity shift as given
by our orbital model. The two spectra were added up using a light ratio of 0.55/0.45 for the primary over the secondary.
The combined theoretical spectra were then compared to the observed spectrum, and the 
$\chi^2$ difference between the two was determined.
For the weights in the $\chi^2$ calculation
we used the inverse square of the S/N for the spectral line.
The total $\chi^2$ for a given combination is the sum of the $\chi^2$ over all observed spectra.
From the minimum $\chi^2$, we found the best fit solution.
As a final verification, we checked all the fits by eye. The best-fit results for a selected number of phases
are shown in Fig.~\ref{all fits}.

While the above procedure uses the 0.55/0.45 light ratio proposed by \cite{Putkuri+23}, we also tried alternative
values. We then judged by eye which gave the better fit. We took into account that
a different light ratio can result in a different best fit combination of stellar parameters.
Based on this, we found that the range 0.525--0.575/0.475--0.425 gives acceptable results.
This conclusion is mainly based on the \ion{He}{I} 4471 and 5876 lines which are most sensitive to
the change in light ratio.

The final results for the effective temperature and gravity of the best-fit TLUSTY models are listed in Table~\ref{table stellar parameters}.
We then used the extinction-corrected value for the V magnitude of the combined system derived by \cite{Putkuri+23}: $m_V=5.283\pm0.022$.
Inverting the Gaia parallax of $0.4985 \pm 0.0204$~mas (\object{HD~168112} = Gaia DR3 4153666106124493696), we have a distance
of $2.01 \pm 0.08$ kpc, which then gave an absolute
$M_V=6.229\pm0.092$. Using the 0.525--0.575/0.475--0.425 light ratio we split this into 
$M_{V,A}=-5.580\pm0.104$ and $M_{V,B}=-5.362\pm0.110$. From the TLUSTY models, we could derive the bolometric correction, and thence
the bolometric luminosity. We could then derive the spectroscopic values for radius and mass.
All stellar parameters are listed in Table~\ref{table stellar parameters}.

\begin{table*}
	\caption{Radio observations of HD 168112.}
	\label{table VLA fluxes}
	\begin{tabular}{llrlllrlll}
		\hline\hline
		\multicolumn{1}{c}{(1)} & \multicolumn{1}{c}{(2)} & \multicolumn{1}{c}{(3)} & \multicolumn{1}{c}{(4)} &
		\multicolumn{1}{c}{(5)} & \multicolumn{1}{c}{(6)} & \multicolumn{1}{c}{(7)} & \multicolumn{1}{c}{(8)} &
		\multicolumn{1}{c}{(9)} \\
		\multicolumn{1}{c}{date} & \multicolumn{1}{c}{BJD} & \multicolumn{1}{c}{band} & \multicolumn{1}{c}{flux (Jy)} & \multicolumn{1}{c}{beamsize} & \multicolumn{1}{c}{PA} & \multicolumn{1}{c}{rms} & \multicolumn{1}{c}{flux} & \multicolumn{1}{c}{$\alpha$} \\
		\multicolumn{1}{c}{time} &  \multicolumn{1}{c}{$-2\,400\,000$}  & & \multicolumn{1}{c}{J1832-1035} & \multicolumn{1}{c}{(\arcsec)} & \multicolumn{1}{c}{($\deg$)} & \multicolumn{1}{c}{(mJy)} & \multicolumn{1}{c}{(mJy)} \\
		\hline	
		2023-Mar-03 & 60\,007.084 & 3.6 cm & $1.407\pm0.002$ & $0.85\times 0.57$ & $-11.46$ & $0.010$ & $0.245\pm0.017$ \\
		13:26:12.0 -- 14:18:57.0 & 60\,007.068 & 6 cm   & $1.400\pm0.005$ & $1.47\times 0.93$ & $ -9.36$ & $0.007$ & $0.130\pm0.011$ & \raisebox{1.5ex}[0pt]{$1.2\pm0.2$}\\
		2023-Mar-18 & 60\,021.994 & 3.6 cm & $1.637\pm0.002$ & $0.98\times 0.64$ & $-30.80$ & $0.035$ & $0.160\pm0.044$ \\
		11:16:33.0 -- 12:09:15.0 & 60\,021.978 & 6 cm   & $1.518\pm0.006$ & $1.95\times 1.09$ & $-31.09$ & $0.017$ & $0.111\pm0.021$ & \raisebox{1.5ex}[0pt]{$0.7\pm0.7$} \\
		2023-Apr-05 & 60\,039.906 & 3.6 cm & $1.436\pm0.001$ & $1.29\times 0.55$ & $-40.76$ & $0.009$ & $0.406\pm0.025$ \\
		09:10:09.0 -- 10:02:54.0 & 60\,039.891 & 6 cm   & $1.475\pm0.007$ & $2.34\times 0.86$ & $-39.36$ & $0.010$ & $0.249\pm0.017$ & \raisebox{1.5ex}[0pt]{$1.0\pm0.2$}\\
		\hline
	\end{tabular}
\tablefoot{
Every two lines in column (1) give the date and time-range of the observation,
column (2) lists the barycentric Julian Date (minus 2\,400\,000),
(3) the wavelength of the observation,
(4) the flux of the phase calibrator,
(5) the size of the synthesized beam, and
(6) its position angle,
(7) the noise near the centre of the image,
(8) the measured flux of HD 168112, and
(9) the spectral index.
}
\end{table*}

Our results present various challenges. A first one is the difference of our stellar parameters (derived from TLUSTY models) 
compared to those 
of \citet[][who use the FASTWIND code]{Putkuri+23}. For the A component, our error bars on $T_{\rm eff}$ and $\log g$ overlap with those 
of \citet[][$T_{\rm eff}=41\,700 \pm 1200$ K, $\log g=3.77 \pm 0.12$]{Putkuri+23}.
However, this is not the case for the B component, where our values are lower than theirs
($T_{\rm eff}=40\,500 \pm 800$ K, $\log g=3.6-3.8 \pm 0.10$). 
A possible cause for this is the difference
between the plane-parallel hydrostatic TLUSTY code and the FASTWIND code that treats in a unified way both the atmosphere and the
(hydrodynamical) stellar wind in spherical symmetry. Another contributing factor is that our procedure was not
applied to the disentangled spectra, but worked directly on the observed data. The disentangled spectra can contain artefacts,
especially in the wings of the hydrogen lines, which are well known to influence the gravity determination.

For the absolute magnitude $M_V$, we achieved sharper error bars, because we relied on the light ratio determined above, rather than the measured
magnitude difference of $0.17\pm0.19$ in H-band \citep{Sana+14}.  
For the spectroscopic bolometric luminosity, we find results that are compatible with the 
\citet[][$\log L/L_{\rm \sun,A} = 5.64 \pm 0.12$, $\log L/L_{\rm \sun,B} = 5.53 \pm 0.08$]{Putkuri+23} ones within
the error bars. Our radii are slightly higher.

Another challenge is that the masses we derived are not nearly equal, in contradiction to the
mass ratio of $0.996^{+0.105}_{-0.183}$ we found
from the orbit determination. 
Comparing our values with the theoretically expected ones from \cite{Martins+05}, we find that our A component mass
is too high (by at least 15\%), while the B component is at least 30\% too low. 
A comparison with \cite{Putkuri+23} shows that our B component mass is compatible
with their result ($M_B < 26$ M$_\sun$), but our A component mass is higher
(although we have marginally overlapping error bars with their result of $M_A=35.4 \pm 8.2$ M$_\sun$). Within the error bars of the
\cite{Putkuri+23} mass for the A component and our value for the B component, we can indeed arrive at a ratio close to one.
Much better limits on the masses of both components can be obtained by combining the spectroscopic data with
astrometric information, as is discussed in Sect.~\ref{section other data}.

\section{Analysis of radio data}
\label{section radio data}

Calibration of the radio data was done by NRAO with the
VLA CASA (Common Astronomy Software Applications) Calibration 
Pipeline\footnote{\url{https://science.nrao.edu/facilities/vla/data-processing/pipeline}}
6.5.4-9.
This pipeline 
flagged data affected by Radio Frequency Interference, and used the calibrator observations to calibrate the \object{HD~168112}
visibilities.
It also applied the flux and band-pass calibration using the known flux of \object{3C286} = J1331+3030. 
We checked by eye the resulting flagging and calibration, and applied some further (light) flagging to the data, as needed.

We then used CASA \citep{CASA} version 6.5.6.22 to convert the \object{HD~168112} calibrated data into images and to deconvolve them, using the {\tt tclean} command.
During this imaging process, we combined all spectral channels, leading to a single image for each of the two bands. We did this to
optimize the S/N. The image fully covers the primary beam, and uses a pixel size of $0\farcs10\times0\farcs10$ for 3.6 cm and $0\farcs15\times0\farcs15$ for 6 cm, which oversamples the synthesized beam well. Towards the edge of the 6 cm image, the resolved image
of a radio galaxy is seen; for this reason we limited the uv-range of the visibility data we used, and we applied multi-scale cleaning. 
We used the \cite{Briggs95} weighting scheme with a 0.5 robust parameter.

We next used the {\tt imfit} command to fit a single elliptical Gaussian to a small area around the centre of the cleaned
image. The integrated flux from this fit and its error bar are listed in Table~\ref{table VLA fluxes}. In the error bar
we also included a 5\% absolute calibration 
error\footnote{\url{https://science.nrao.edu/facilities/vla/docs/manuals/oss/performance/fdscale}}
which has been added in quadrature to the measured flux error.
To judge the systematic errors, we repeated the reduction a number of times, dropping data from one of the antennas,
or one of the three time sequences of the observations. In all
cases, the results fall within the error bar listed in Table~\ref{table VLA fluxes}. 

From the fluxes at two wavelengths, we derived the spectral index $\alpha$, which is defined by $F_\nu \propto \nu^\alpha$. 
This is also listed in Table~\ref{table VLA fluxes}. The interpretation of these data is discussed in the next section.

\section{Comparison with other data}
\label{section other data}

Radio and optical interferometry data of \object{HD~168112} have been published in the literature. In this section, we 
compare our results to those data.

\cite{Blomme+05} analysed the available archive radio data on \object{HD~168112} from the Very Large Array
(VLA) and from the Australia Telescope Compact Array (ATCA), They found considerable variability in the fluxes, but
the radio data did not allow for a precise determination of the orbital period. A value
between one and two years was proposed, with a most likely value of ~$\sim 1.4$ yr. The latter is in good
agreement with our value of 512.17 days.

\begin{figure}
	\resizebox{\hsize}{!}{\includegraphics[bb=70 360 340 790]{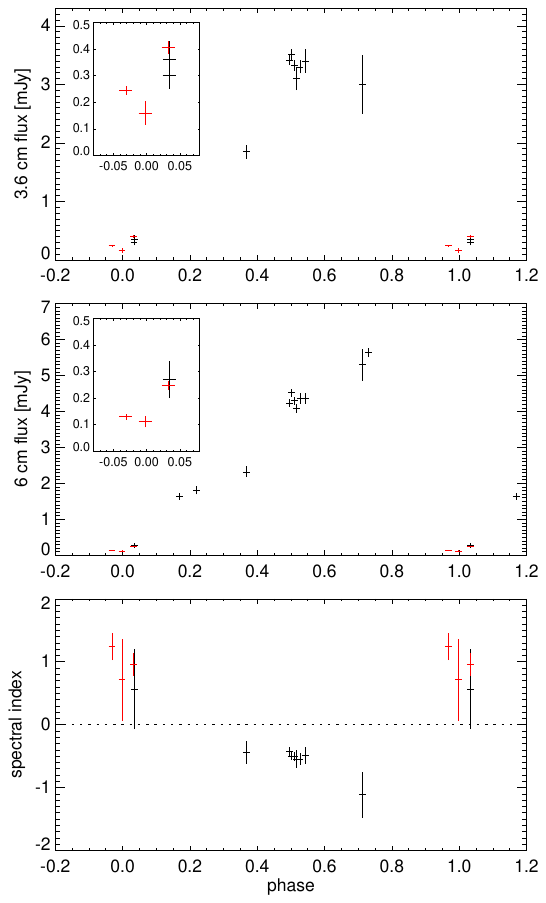}}
	\caption{Radio fluxes at 3.6 cm (\textit{top panel}) and 6 cm (\textit{middle panel}),
		plotted as a function of the orbital phase in the 512.17 d period,
		and the corresponding spectral index (\textit{bottom panel}). The black symbols show the data from 
		\cite{Blomme+05} while the red ones give the data presented in this paper. The insets zoom in on
		the phase range around periastron.
	}
	\label{fig radio data}
\end{figure}

Now that the orbital parameters are known, we can plot again the radio fluxes of \cite{Blomme+05} in the orbital
phase diagram (Fig.~\ref{fig radio data}), and add the new observations. 
The radio fluxes show a clear correlation with the orbital phase. The fluxes
are at minimum near periastron, and the new observations presented here (in red on Fig.~\ref{fig radio data})
are in excellent agreement with the older data, which were taken $\sim$~20--40 years before. 

The spectral index (Fig.~\ref{fig radio data}, bottom plot) allows us to distinguish between thermal and non-thermal radiation.
Thermal radiation is due to free-free emission in the stellar winds of both stars and has a spectral index 
of $\alpha \approx +0.6$ \citep{Wright+Barlow75, Panagia+Felli75}. 
Where the two stellar winds collide, shocks form on either side of the collision
region. Around these shocks, a fraction of electrons can attain relativistic speeds due to
the Fermi mechanism \citep{Bell78,Reitberger+14, Pittard+21}. By spiralling around in the magnetic
field, these electrons then emit synchrotron radiation, which we detect as non-thermal radiation 
\citep{Eichler+Usov93}. Theoretical models show that the non-thermal spectral index can be almost as low as $-1.0$
\citep{Pittard+06}. 

As had already been noted by \cite{Blomme+05}, the radio fluxes of \object{HD~168112} clearly show a non-thermal spectral index, at least
at orbital phases away from periastron. Near periastron, however, the spectral index is more consistent with thermal
emission. 
We can first estimate the expected thermal flux from both stellar winds. Using the stellar parameters from Table~\ref{table stellar parameters},
we applied the \cite{Vink+01} mass-loss recipe to determine the mass-loss rate. 
We found $2.1\times 10^{-6}$ ${\rm M}_\sun {\rm yr}^{-1}$ for component A and
$1.2\times 10^{-6}$ ${\rm M}_\sun {\rm yr}^{-1}$ for component B.
For the terminal velocity, we used the
value of 3250 km\,s$^{-1}$ from \cite{Leitherer88}. We derived the distance from
the Gaia parallax.
The \cite{Wright+Barlow75} formula for the theoretical radio flux then gives 0.029 mJy (component A) + 0.014~mJy (component B)
at 3.6 cm and 
0.022 + 0.010 mJy at 6 cm. 
This is at least a factor 3--4 lower than the observed fluxes near periastron. 
Clumping in the stellar wind increases the radio flux for a given mass-loss rate.
The \cite{Vink+01} mass-loss rates do not include clumping, but recent work by, for example,
\cite{Hawcroft+21} on spectroscopic data shows that the inclusion of clumping requires a decrease in the mass-loss
rate \citep[compared to][]{Vink+01} in order to explain the strengths of the spectral lines.
To first order, the effect of the clumping and the reduction of the mass-loss rate compensate one another.
We therefore cannot attribute the observed radio fluxes near periastron to clumping in the winds.
It is therefore probable that the wind-wind
collision region is also contributing thermal radiation, as the theoretical models of \cite{Pittard10} predict.

In this highly eccentric binary, one expects the stellar wind collision to be strongest
at periastron and therefore the highest synchrotron flux to be generated. 
The question therefore remains why we do not detect this synchrotron radiation.
One possible reason is that the shocks collapse at periastron.
However, this is contradicted by the X-ray flux variations that are symmetrical around the periastron passage
\citep{Rauw+24}.
Another possibility is that 
the synchrotron photons are being absorbed by the free-free absorption in the two stellar winds. 
From Table~\ref{table orbital parameters}, we find that at periastron, the two components
have a projected separation of 220 R$_\sun$. This should be compared to the typical size of the radio emitting region,
for which we use the radius where the optical depth $\tau_\lambda=1$. Using the \cite{Wright+Barlow75} equations
and the mass-loss rate and terminal velocity determined above, we find a 3.6 cm radius of 420 R$_\sun$ for component
A and 300 R$_\sun$ for component B. At 6 cm, the values are a factor $(6/3.6)^{2/3} \approx 1.41$ higher. 
The introduction of clumping has little influence on these values because, again, to first order,
the effect of the clumping and the reduction of the mass-loss rate compensate one another.
At periastron the synchrotron emission region is therefore well inside the two stellar winds.
It is only when the stars
are further apart in their orbit 
(for example, at apastron, the projected separation is 1570 R$_\sun$)
that a much higher fraction of the synchrotron photons can escape and be detected.
This effect has also been seen in other binaries, such as \object{WR 140} \citep{WR140-1,WR140-2} and 
\object{Cyg OB2 \#9} \citep{VanLoo+08}.

\cite{DeBecker+24} observed \object{HD~168112} at 1.6 GHz ($18$~cm) with the European Very Long Baseline Interferometry Network (EVN). 
They obtain a high spatial resolution image that resolves the colliding-wind region. The region
is clearly asymmetric and its high flux (integrated $1.7\pm0.4$~mJy) shows it to be non-thermal in origin. The date of
their observation (2019-Nov-05) corresponds to phase 0.60 according to our orbital solution.
A direct comparison with the results presented here is not possible, as the orbital phase and wavelength of the observation are quite
different. \cite{Blomme+05} listed a much higher 20 cm flux of $11\pm7$ mJy at phase 0.68 (project AC116, date 1984-Nov-27)
and $11\pm2$ mJy at phase 0.71 (C978, 2001-Oct-11). However, these fluxes were derived from VLA data, which do not resolve the
colliding-wind region.
As \cite{DeBecker+24} show, the radio fluxes are dominated by the colliding-wind region and the contribution of the 
two stellar winds is negligible. The higher VLA fluxes compared to the EVN ones therefore show that 
the synchrotron emitting region must be geometrically more extended than
is shown in the EVN image. In the outer parts of this extended region, the intensity is lower than in the
inner parts, and so the outer parts are
not detected on the EVN image. However, their large geometric extent makes them contribute substantially to the total flux.

\cite{Sana+14} found a visual companion to \object{HD~168112} using the VLTI PIONEER instrument. The companion was detected
at a distance of $3.33 \pm 0.17$~mas at a position angle of $303\fdg27 \pm 4\fdg12$. The magnitude difference in
the H band is $0.17 \pm 0.19$. The date of the observation is 2012.4446, which corresponds to JD 2\,456\,091.224. Using
our orbital solution we found this corresponds to phase 0.32. At that phase the radius
divided by the semi-major axis is $r_A/a_A \approx r_B/a_B = 1.6$. 

We can convert the angular separation of $3.33$~mas to a projected linear distance of $\sim 1436$~R$_\sun$
using the Gaia parallax. To compare this with our spectroscopic orbit, we need to use the following astrometric
conversion \citep{Gallenne+23}:
\begin{eqnarray}
\Delta \alpha & = & r [\sin \Omega \cos(\omega+\nu) + \cos i \cos \Omega \sin(\omega+\nu)], \nonumber \\
\Delta \delta & = & r [\cos \Omega \cos(\omega+\nu) - \cos i \sin \Omega \sin(\omega+\nu)], \label{eq astrometric}\\
r & = & \frac{a(1-e^2)}{1+e\cos\nu}, \nonumber
\end{eqnarray}
\noindent
where $\Delta \alpha$ and $\Delta \delta$ are the offset in right ascension and declination, $r$ is the (non-projected) separation,
$\Omega$ is the longitude of the ascending node, $\nu$ is the true anomaly, and the other parameters have been defined in Sect.~\ref{section preliminary orbit}. The projected separation is then given by:
\begin{eqnarray}
\rho & = &\sqrt{(\Delta \alpha)^2 + (\Delta \delta)^2} \nonumber\\
     & = & \frac{a(1-e^2)}{1+e\cos\nu} \sqrt{\cos^2(\omega+\nu)+\cos^2i\sin^2(\omega+\nu)}. \nonumber
\end{eqnarray}
From the spectroscopic value of $a \sin i$ and the astrometrically measured $\rho$, we then find the inclination to be $\sim$~63\degr.

Using this value for the inclination and the minimum masses from Table~\ref{table orbital parameters}, 
we then find the masses to be $\sim 26~{\rm M}_\sun$.
This is compatible with the spectroscopic mass derived for the B component, but too small to be consistent
with the A component (Table~\ref{table stellar parameters}). 

The position angle $\theta$ (measured from the north to the east) can be derived from:
\begin{equation}
\theta = \frac{\pi}{2} - \arctan\left(\frac{\Delta \delta}{\Delta \alpha}\right), \nonumber \\
\end{equation}
\noindent
where care must be taken to assign $\theta$ to the correct quadrant.
From Eqs.~\ref{eq astrometric}, we can also derive $\Omega$, although there is an ambiguity as we do not know if
the measured position angle is from component A to B or vice versa. Assuming the $303\fdg27$ position angle is from A to B, we
find $\Omega\approx 318\degr$, otherwise $\Omega\approx 138\degr$.

The \cite{DeBecker+24} resolved radio observation also provides an estimate of the 
position angle. On their Fig.~1, the colliding wind region is asymmetric with the west side wrapping around the
star with the weakest wind (which we identify with the B component). From this we can estimate an A to B position
angle\footnote{Our value is different from the one they list, as they measure the position angle starting from the east.} of
$\theta \approx 298\degr$. From this we find $\Omega \approx 301\degr$, which is compatible with one of the results from the
optical interferometry.

\section{Conclusions}
\label{section conclusions}

We analysed 113 spectroscopic observations of HD~168112 taken during a 13 year time span with 
the HERMES and TIGRE spectrographs. In this way, we obtained an orbital solution
that is independent of the one determined by \cite{Putkuri+23}. We find a period
of $512.17^{+0.41}_{-0.11}$~d, an eccentricity of $0.7533^{+0.0053}_{-0.0124}$,
and a mass ratio close to one.
A comparison
with the \cite{Putkuri+23} results shows that we have a slightly shorter period. To determine
the radial velocities for each observation, we used an iterative approach with
the disentangled spectra providing the shape of each spectral line. This
approach differs from the \cite{Putkuri+23} one, who used Gaussian profiles.
As a consequence, we find slightly smaller semi-amplitudes for both components,
and therefore also smaller lower limits on the masses.

We also derived the stellar parameters for the two components. Contrary to \cite{Putkuri+23}
we did not use
the disentangled spectra for this, but instead tried various combinations of 
two model spectra and compared the combined spectrum (with the appropriate
radial velocity shifts) to the observed spectra at various orbital phases.
 Within the (large) error bars we find
similar $T_{\rm eff}$ and $\log g$ for the A component, but our B-component
values are lower than the \cite{Putkuri+23} ones. The difference could be 
due to the different procedure we used (avoiding in our case the known disentangling problems), or to 
our use of plane-parallel TLUSTY models instead of the \cite{Putkuri+23} FASTWIND ones.

We also analysed radio observations that were taken at three epochs close
to the periastron passage. We expect the synchrotron emission around the colliding wind
region to result in a negative spectral index. 
But the spectral index of all three observations shows only 
thermal emission, suggesting that most of the synchrotron photons are absorbed
in the two stellar winds. However, the total flux cannot be attributed to the stellar winds only.
The colliding-wind region must therefore also be contributing thermal radio emission.
Clearly, this effect can also occur in other colliding-wind binaries.

By combining our data with the optical interferometry result of
\cite{Sana+14}, we can constrain the inclination angle ($i \sim 63\degr$), 
and in this way estimate the masses of both components ($\sim 26$ M$_\sun$). 
For the A component, this is considerably lower than the spectroscopic estimate,
while the value is consistent with the B component.
As the error bars on the orbital parameters are relatively small, we have a higher
confidence in the masses derived in this way, rather than the spectroscopic masses.
Finally, we showed that our orbital parameters are
consistent with the resolved radio image obtained by \cite{DeBecker+24}.

\begin{acknowledgements}
We are grateful to the observers at the Mercator Telescope
for collecting the data of our HERMES service mode programme.
The Li\`ege team acknowledges support from the Fonds 
National de la Recherche Scientifique (Belgium) and the Belgian Federal Science
Policy Office (BELSPO) in the framework of the PRODEX Programme (contract
HERMeS).
This work has used the following software products:
emcee \citep{emcee},
corner.py \citep{corner},
Matplotlib \citep[][\url{https://matplotlib.org}]{Hunter:2007},
NumPy \citep[][\url{https://numpy.org}]{harris2020array},
IDL (\url{https://www.nv5geospatialsoftware.com/Products/IDL}),
\cite{Shenar+20,Shenar+22} disentangling code,
{\sc iacob-broad} code \citep{IACOB_BROAD},
CASA \citep{CASA}.
This work has made use of data from the European Space Agency (ESA) mission
{\it Gaia} (\url{https://www.cosmos.esa.int/gaia}), processed by the {\it Gaia}
Data Processing and Analysis Consortium (DPAC,
\url{https://www.cosmos.esa.int/web/gaia/dpac/consortium}). Funding for the DPAC
has been provided by national institutions, in particular the institutions
participating in the {\it Gaia} Multilateral Agreement.
This research has made use of the SIMBAD database, operated at CDS, Strasbourg, France, and
NASA’s Astrophysics Data System.
We thank the National Radio Astronomy Observatory (NRAO) for carrying out
the Karl G. Jansky Very Large Array (VLA) observations and applying the CASA pipeline reduction.
We thank the referee for comments that helped clarify the paper.
\end{acknowledgements}

\bibliographystyle{aa}
\bibliography{hd168112}

\begin{appendix}
	\onecolumn
\section{Observing log}

\begin{table}[h!]
	\caption{Observing log of the HD~168112 spectra.}
	\label{table:obslog}
	\centering
	\begin{tabular}{rrrrr|rrrrrrr}
		\hline\hline
		\multicolumn{1}{c}{(1)} & \multicolumn{1}{c}{(2)} & \multicolumn{1}{c}{(3)} & \multicolumn{1}{c}{(4)} &&& \multicolumn{1}{c}{(1)} & \multicolumn{1}{c}{(2)} & \multicolumn{1}{c}{(3)} & \multicolumn{1}{c}{(4)} \\
		\multicolumn{1}{c}{file} & \multicolumn{1}{c}{BJD} & \multicolumn{1}{c}{exp.time} & \multicolumn{1}{c}{S/N} & &&
	    \multicolumn{1}{c}{file} & \multicolumn{1}{c}{BJD} & \multicolumn{1}{c}{exp.time} & \multicolumn{1}{c}{S/N} \\
	    & \multicolumn{1}{c}{-2\,400\,000} & \multicolumn{1}{c}{(s)} & & & & & \multicolumn{1}{c}{-2\,400\,000} & \multicolumn{1}{c}{(s)} & \\
\hline
\multicolumn{3}{c}{HERMES} & & & & \multicolumn{3}{c}{HERMES}\\
00236514 & 54995.6722 &  2700 &   211  & && 00724589 & 57590.5491 &  1800 &    84 \\
00236515 & 54995.7047 &  2700 &   153  & && 00724731 & 57591.4126 &  1800 &   101 \\
00240524 & 55017.5159 &  1800 &   198  & && 00724732 & 57591.4354 &  1800 &   109 \\
00240529 & 55017.5691 &  1800 &   174  & && 00727747 & 57612.4945 &  1800 &   106 \\
00240530 & 55017.5907 &  1800 &   191  & && 00728994 & 57623.3961 &  1800 &   116 \\
00242124 & 55036.5373 &  1800 &   131  & && 00777064 & 57863.6286 &  1800 &   126 \\
00242125 & 55036.5591 &  1800 &   121  & && 00777069 & 57863.7035 &  1800 &   143 \\
00242126 & 55036.5805 &  1800 &   115  & && 00777204 & 57864.6340 &  1800 &   147 \\
00247264 & 55085.3975 &  2700 &   201  & && 00777210 & 57864.7067 &  1800 &   148 \\
00247265 & 55085.4294 &  2700 &   194  & && 00782516 & 57892.5725 &  1800 &   125 \\
00288714 & 55373.4851 &  1800 &   170  & && 00782522 & 57892.7036 &  1800 &   125 \\
00288715 & 55373.5066 &  1800 &   162  & && 00782652 & 57893.6243 &  1800 &   160 \\
00288716 & 55373.5280 &  1800 &   177  & && 00785624 & 57900.5626 &  1800 &   163 \\
00352807 & 55708.6270 &  1800 &   111  & && 00785757 & 57901.6855 &  1800 &   159 \\
00352808 & 55708.6484 &  1800 &   107  & && 00785758 & 57901.7079 &  1800 &   161 \\
00352809 & 55708.6698 &  1800 &   105  & && 00787025 & 57910.6291 &  1800 &    74 \\
00361488 & 55769.4354 &  1800 &    60  & && 00787151 & 57911.5127 &  1800 &    74 \\
00361489 & 55769.4569 &  1800 &    66  & && 00793320 & 57919.5886 &  1800 &   107 \\
00361490 & 55769.4783 &  1800 &    73  & && 00819759 & 57935.4803 &  1800 &   105 \\
00361491 & 55769.5012 &  1800 &    86  & && 00822228 & 57938.5590 &  1800 &   146 \\
00361590 & 55770.4713 &  1800 &   150  & && 00822345 & 57939.5313 &  1800 &   141 \\
00361591 & 55770.4927 &  1800 &   137  & && 00834212 & 57977.4673 &  2400 &   131 \\
00361592 & 55770.5141 &  1800 &   145  & && 00834215 & 57977.5224 &  2400 &   121 \\
00412855 & 56112.5262 &  1800 &   167  & && 00834349 & 57978.4695 &  2100 &   140 \\
00412856 & 56112.5477 &  1800 &   172  & && 00835205 & 57984.4783 &  1800 &   134 \\
00412857 & 56112.5691 &  1800 &   170  & && 00835336 & 57985.4713 &  1800 &   118 \\
00415761 & 56141.5526 &  1800 &   123  & && 00835467 & 57986.4698 &  1800 &    94 \\
00476112 & 56462.5220 &  1800 &   148  & && 00835601 & 57987.4229 &  1800 &    95 \\
00476113 & 56462.5434 &  1800 &   148  & && 00835602 & 57987.4455 &  1800 &    79 \\
00476114 & 56462.5648 &  1800 &   157  & && 00835830 & 57989.4662 &  1800 &   121 \\
00562316 & 56813.5847 &  1800 &   151  & && 00844339 & 58016.3353 &  1800 &   100 \\
00562317 & 56813.6061 &  1800 &   160  & && 00844340 & 58016.3577 &  1800 &   100 \\
00562318 & 56813.6275 &  1800 &   154  & && 00963844 & 58994.5967 &  1800 &   176 \\
00573490 & 56829.5528 &  1800 &   157  & && 00964060 & 58996.5822 &  1800 &   194 \\
00573491 & 56829.5742 &  1800 &   157  & && 00964293 & 58998.6052 &  1800 &    92 \\
00573492 & 56829.5957 &  1800 &   151  & && 00968698 & 59044.5499 &  1800 &   173 \\
00574707 & 56846.4778 &  1800 &   156  & && 00974649 & 59101.3725 &  1900 &   140 \\
00574708 & 56846.4992 &  1800 &   162  & && 01013427 & 59452.4973 &  1910 &   146 \\
00574709 & 56846.5206 &  1800 &   156  & && 01013961 & 59457.3839 &  1800 &   151 \\
00578766 & 56878.4582 &  1800 &   190  & && 01015103 & 59468.3436 &  1800 &   131 \\
00578767 & 56878.4796 &  1800 &    82  & && \multicolumn{3}{c}{TIGRE} & & &\\
00579276 & 56882.4391 &  1800 &   183  & && T20170819 & 57985.6955 &  1800 &    27 \\
00579277 & 56882.4605 &  1800 &   180  & && T20170823 & 57989.6940 &  1800 &    27 \\
00579278 & 56882.4819 &  1800 &   172  & && T20170902 & 57999.5888 &  1800 &    39 \\
00646356 & 57197.5525 &  2050 &   185  & && T20170907 & 58004.6669 &  1800 &    22 \\
00646357 & 57197.5768 &  2050 &   175  & && T20170911 & 58008.6700 &  1800 &    35 \\
00646358 & 57197.6012 &  2065 &   176  & && T20170915 & 58012.6663 &  1800 &    39 \\
00651693 & 57237.3949 &  1900 &   156  & && T20200425 & 58964.8366 &  3600 &    45 \\
00651694 & 57237.4174 &  1900 &   165  & && T20200429 & 58968.8337 &  3600 &    42 \\
00651695 & 57237.4400 &  1900 &   156  & && T20200503 & 58972.9090 &  3600 &    59 \\
00711066 & 57514.6095 &  2500 &   129  & && T20200508 & 58977.9538 &  3600 &    32 \\
00717344 & 57552.5783 &  1800 &   117  & && T20200512 & 58981.8719 &  3600 &    47 \\
00717345 & 57552.6008 &  1800 &   118  & && T20200516 & 58985.8691 &  3600 &    49 \\
00717346 & 57552.6232 &  1800 &   124  & && T20200520 & 58989.8663 &  3600 &    45 \\
00719032 & 57564.4981 &  2000 &    80  & && T20200524 & 58993.8030 &  3600 &    46 \\
00719169 & 57565.5591 &  1800 &    96  & && T20200530 & 58999.8539 &  3600 &    33 \\
00724588 & 57590.5350 &   284 &    28  & && T20200606 & 59006.8514 &  3600 &    35 \\
\hline
    \end{tabular}
\tablefoot{
Column (1) lists file number, (2) barycentric Julian Date of middle of observation, minus 2\,400\,000, (3) exposure time, (4) S/N at 5410 \AA.
}
\end{table}

\begin{figure}[ht]
	\centering
	\includegraphics[width=12cm,viewport=30 30 505 430,clip]{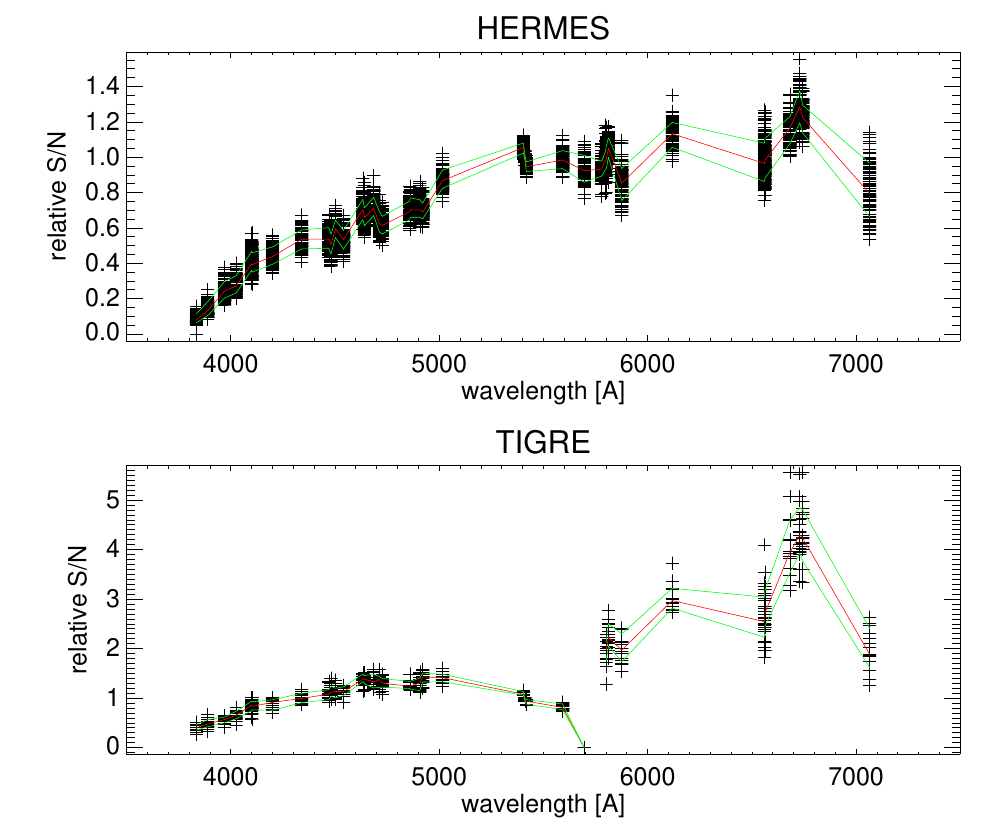}
	\caption{Relative signal-to-noise ratio (S/N) as a function of the wavelength for the HERMES ({\em top panel})
		and the TIGRE ({\em bottom panel}) spectra. The S/N is relative to the S/N at 5410 \AA, with the
		5410 \AA~value being listed in Table~\ref{table:obslog}. At a given wavelength, the multiple crosses
		show the value of the individual spectra for the spectral line. The red line gives the median
		of these values, and the green lines the range that contains 68.3\% of the data.
		The break in the TIGRE data around 5700 \AA\ is due to the change from the blue arm
		to the red arm \citep{TIGRE}.}
	\label{fig S/N}
\end{figure}
\clearpage
\section{Spectral line fitting}

\begin{figure}[ht]
	\centering
    \includegraphics[width=15.6cm,viewport=22 20 560 770,clip]{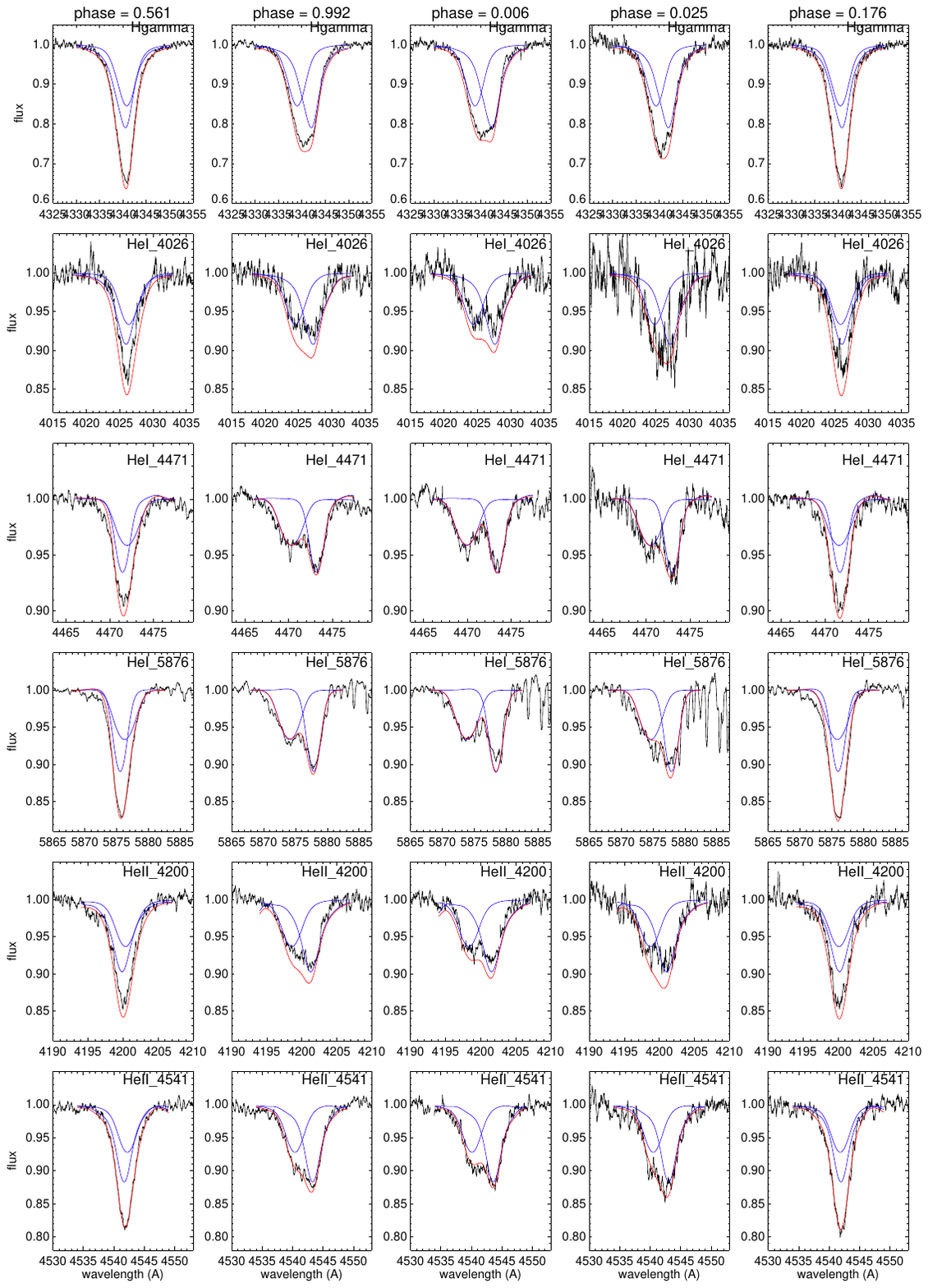}
	\caption{Best fit to the spectral lines for a selection of orbital phases, using the stellar parameters from
    Table~\ref{table stellar parameters}. Each column shows an orbital phase
	(listed at the top), each row shows a spectral line (listed inside each plot). The observed spectrum is given by the black curve, the theoretical
    fits of each of the two components by the blue curves, and the combined fit by the red curve.}
	\label{all fits}
\end{figure}

\end{appendix}

\end{document}